\documentclass{aastex}
\usepackage{amsmath}
\usepackage{lscape,graphicx}
\usepackage{epsfig}
\usepackage{verbatim}
\usepackage{caption}

\shorttitle{Supergranular intensity profile}
\shortauthors{Goldbaum {\it et al.}}

\begin{document}

\title{The Intensity Profile of the Solar Supergranulation}

\author{Nathan Goldbaum}
\affil{Laboratory for Atmospheric and Space Physics and Department of Physics, University of Colorado, Boulder, CO 80309}
\author{Mark P. Rast\altaffilmark{1}}
\altaffiltext{1}{Corresponding author: mark.rast@lasp.colorado.edu}
\affil{Laboratory for Atmospheric and Space Physics, Department of Astrophysical and
Planetary Sciences, University of Colorado, Boulder, CO 80309 USA}
\affil{and}
\affil{High Altitude Observatory, National Center for Atmospheric Research\altaffilmark{2}, PO Box 3000, Boulder, CO 80307, USA}
\author{Ilaria Ermolli}
\affil{INAF - Osservatorio Astronomico di Roma, Via di Frascati, 33, I-00040 Monte Porzio, Italy}
\author{J. Summer Sands}
\affil{SOARS and the High Altitude Observatory, National Center for Atmospheric Research\altaffilmark{2}, PO Box 3000, Boulder, CO 80307, USA}
\altaffiltext{2}{NCAR is sponsored by the National Science Foundation.}
\author{Francesco Berrilli}
\affil{Dipartimento di Fisica, Universit$\rm{ \grave a}$ di Roma Tor Vergata, Viale della Ricerca Scientifica, 1, I-00133 Roma, Italy}

\begin{abstract}
We have measured the average radial (cell center to network boundary) profile of the
continuum intensity 
contrast associated with supergranular flows using data from the 
Precision Solar Photometric Telescope (PSPT) at the Mauna Loa Solar Observatory (MLSO).
After removing the contribution of the network flux elements by the application of masks based on Ca II K intensity and averaging over more than $10^5$ supergranular cells, we find a $\sim 0.1\%$ decrease in red and blue continuum intensity from the supergranular cell centers outward, corresponding to a $\sim 1.0$ K decrease in brightness temperature across the cells.  
The radial intensity profile may be caused either by the thermal signal associated with the supergranular flows or a variation in the packing density of unresolved magnetic flux elements.  
These are not unambiguously distinguished by the observations, and  
we raise the possibility that the network magnetic fields play an active role in supergranular 
scale selection by enhancing the radiative cooling of the deep photosphere at the cell boundaries. 
\end{abstract}


\section{INTRODUCTION}

The solar supergranulation was first observed by \citet{har54} as coherent banded velocities structures with  length scales of approximately 75 Mm in line of sight Doppler measurements at the Sun's limb.  Subsequent work \citep{lei62,sim64} concluded that the signal observed by Hart corresponds to a surface-filling cellular pattern of horizontally-diverging flows with diameters of approximately 30 Mm, flow speeds of $\sim$ 500 m s$^{-1}$, and lifetimes of $\sim$ 1 day.  The same studies reported the strong statistical correspondence between the chromospheric magnetic network observed in Ca II K images and the borders of the supergranular cells.  This correlation was confirmed by~\citet{kue83}.  

Supergranulation is readily identified in time-averaged Doppler images \citep[e.g.][]{hat00}, by tracking the motions of individual granules \citep[e.g.][]{rie08}, or by using the correlation between the Ca II K network and the convergent flow boundaries \citep[e.g.][]{ber98}.  The characteristic cell sizes reported depend somewhat on the identification technique employed, ranging from 15 Mm \citep{der04} to 25 Mm \citep{hir08} or 35 Mm \citep{rie08}, and may vary with the solar cycle \citep{ber99,meu08}.  Supergranular flows are characterized by $\sim 0.07$ km s$^{-1}$ central upflows \citep{kue83} and well observed horizontal outflows with average speeds of $\sim 0.5$ km s$^{-1}$ \citep{meu08} that converge on isolated sites of downflow at supergranular borders \citep{kue83}.  Measurements of downflow speeds are limited by the intrinsically small downflow velocities as well as the presence of the magnetic network fields  \citep[e.g.][]{fru98}. Taken together, these observations suggest that fluid emerges near the supergranular cell center, flows outward towards the cell borders advecting the magnetic field, before sinking back into the solar interior. It is thus generally inferred that supergranulation is thermally driven and has a convective origin.  

Early on \citep{lei62}, and later reinforced by the discovery of mesogranulation \citep{nov81}, it was sugggested that the somewhat discrete scales of solar convection reflect the depths of hydrogen and helium ionization.  This is not supported by detailed models of ionizing convection, which show that partial ionization is broadly distributed in depth (ionization/recombination occurs at very different depths in upflows and downflows) and causes both linear and finite amplitude destabilization of the flow, favoring high-speed small-scale motions \citep{ras91,ras93,ras01}, not large-scale flows with length scales reflecting the poorly defined mean ionization depth.  This has led to the suggestion that the more elusive mesogranular and the supergranular scales emerge directly from the self-organization of granular flows \citep[e.g.][and reference therein]{rie00,ras03a}.
While unambiguously determining the driving mechanism from observations is likely difficult, measurement of the thermal perturbation associated with supergranular flows, the goal of our work, places constraints on these models. 

Measurement of the radial (cell center to network boundary) profile of the supergranular intensity must overcome two challenges:  the weak supergranular signal underlies intensity fluctuations due to both granulation and the magnetic network.  The granular "noise" can be overcome by spatial and temporal averaging since granulation evolves on short time scales and small length scales compared to those of the supergranulation.  The contribution due to the magnetic network, however, must be more carefully accounted for, since the network is itself closely associated with the supergranular flows.  Absent a correction for the network contribution, the supergranular contrast at most continuum wavelengths peaks at the cell boundaries \citep{bec68,fou84,lin92}.  This is seemingly inconsistent with models for the supergranulation in which warm fluid rises at cell center and cool fluid descends into the solar interior at the network boundaries.  However, these uncorrected contrast measurements are dominated by opacity changes due to the presence of magnetic flux elements \citep[e.g.][]{spr76,piz93,ste05,cri09}, obscuring any thermal effects due to temperature perturbations in the plasma.  Once the magnetic and thermal contributions to the signal have been disentangled, a decrease in brightness from the center to the edge of an average supergranule has been reported \citep{ras03b,meu07,meu08}, corresponding to a center-to-edge temperature change of 0.8 to 2.8 K \citep{meu07,meu08}.  Our work confirms these earlier measurements with a higher degree of certainty and somewhat better spatial resolution, while employing identification and measurement methodologies distinctly different from those of Meunier and coworkers.

\label{intro}
\section{ANALYSIS}

\subsection{The PSPT Data}

The Precision Solar Photometric Telescope (PSPT) is a 15 cm refracting telescope designed for high (0.1\%) pixel-to-pixel photometric precision, operated by the High Altitude Observatory (HAO) at Mauna Loa Solar Observatory (MLSO).  It produces 2048 $\times$ 2048 full-disk solar images ($\sim1" {\rm pixel}^{-1}$) with a $\sim 20$ minute cadence, weather permitting, in five wavelength bands: red continuum (607.1 FWHM 0.5 nm), blue continuum (409.4 FWHM 0.3 nm), Ca II K (393.4 FWHM 0.3 nm), and two narrow-band Ca II K filters not used in this study.  Data processing challenges and techniques are described in detail in \citet{ras08}.  

For this study a total of 1051 image triplets, nearly simultaneous (interleaved during observation) red continuum, blue continuum, and Ca II K images, from the period January 2005 to March 2008 were selected.  
Selection was based on the quality of the red image \citep[the images used in this study have maximum red and blue continuum limb widths of 2.5 and 3.5 pixels respectively, see][]{ras08} and the availability of images in all three wavelength bands.  

Since our final measurement of the average radial intensity profile of supergranular cells makes use of only a small fraction of the pixels in any given image, those remaining after masking out magnetic element contriutions, the signal from many images must be combined into a statistically significant measurement.  To this end, each image triplet is aligned, supergranules are identified in the central $512^2$ pixel region of all 1051 Ca II K images, and the average intensity at each wavelength as a function of distance from the supergranular barycenter is computed, with total of more than $10^5$ supergranules contribute to the final measurement.
\label{instrument}

\subsection{Data Analysis}

The network pattern associated with the supergranular flows is readily apparent by visual inspection of the PSPT Ca II K images, but  unambiguous and automatic identification of this pattern is required to perform statistical analyses of a large ensemble of supergranules.  Two identification methods are commonly reported in the literature.  The first begins with local correlation tracking of granular features to determine the advective flow of the supergranulation \citep[e.g.][]{nov88,pot08}.  The divergence of the horizontal velocity field measured is then calculated and smoothed, and a steepest descent algorithm is applied to compute the cell boundaries \citep{der04,meu07}.  The second, the method employed here, makes use of the strong correlation between the borders of supergranules observed in velocity and the chromospheric network observed in the Ca II K band \citep{kue83}, which in turn reflects the correlation between Ca II K emission and magnetic flux density \citep{sku75,sch89,har99,ras03b,ort05} and the convergent advection of magnetic elements into the supergranular boundaries.  In this method, a computer vision algorithm, based on morphological operations that segment binarized Ca II K images \citep{ber98,ber99,ber05}, maps the connected network boundaries. Mixed methods are certainly feasible, and should be explored for consistency in future work. 

We employ the iterative medial axis transform algorithm of \citet{ber98,ber05} to generate supergranular maps of the central $512^2$ portion of the PSPT Ca II K contrast images\footnote{The contrast at each wavelength is defined as  $I_c\equiv \frac{\delta I}{I_0}=\frac{I-I_0}{I_0}$, where
$I$ is the observed gain-corrected intensity and $I_{0}$ is the measured center-to-limb function at that wavelength \citep{ras08}.}.  The algorithm calculates a binary image by boxcar-averaging the Ca II K contrast image and uses this binary image to perform a skeletonization.  If $I_K(x,y)$ is the Ca II K contrast at pixel position $(x,y)$, then the binary image is that produced by a high pass threshold based on the local mean and standard deviation of the the pixels in an $L \times L$ moving window, with the threshold function $T(x,y)= \langle I_K(x,y) \rangle_L + \xi\sigma(x,y)_L,$ where $\sigma(x,y)_L$ and $\langle I_K(x,y) \rangle_L$ are, respectively, the standard deviation and average intensity of the pixels in the $L \times L$ window centered at $(x,y)$.  $\xi$ is a tunable parameter.  

\begin{figure}
\centering
\includegraphics[scale=.85]{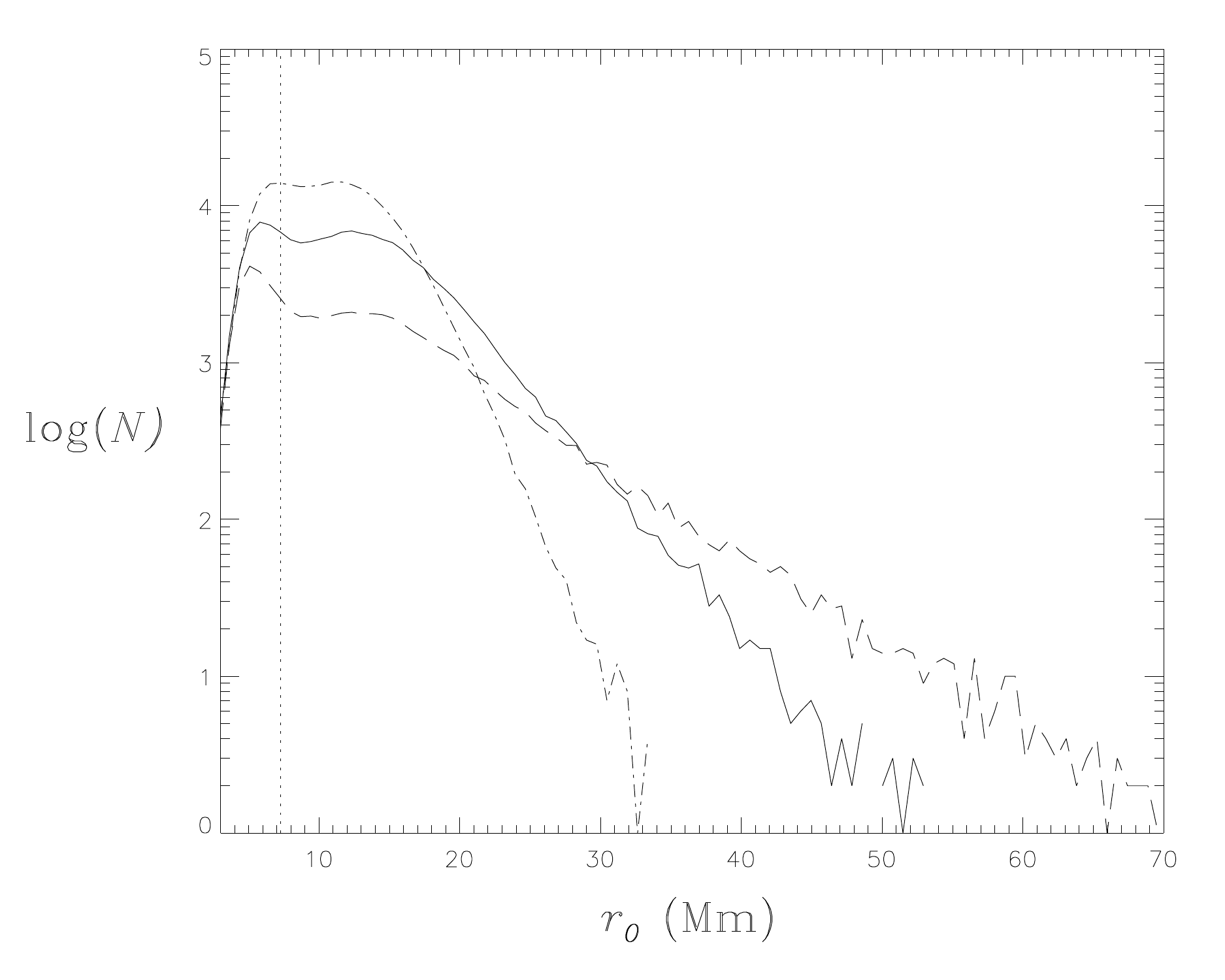}
\caption{(Online) Logarithm of the number of supergranular cells identified in the images as a function of there size for three values of $\xi$. The dashed line corresponds to $\xi = -0.2$, the solid line corresponds to $\xi = -0.4$, and the dot-dashed line corresponds to $\xi = -0.6$.  The dotted fiducial line indicates the length cutoff below which cells are not considered for the annular averaging in the primary analysis.}\label{cellsizehisto}
\end{figure}

\begin{figure}
\begin{center}
\includegraphics[scale=.65]{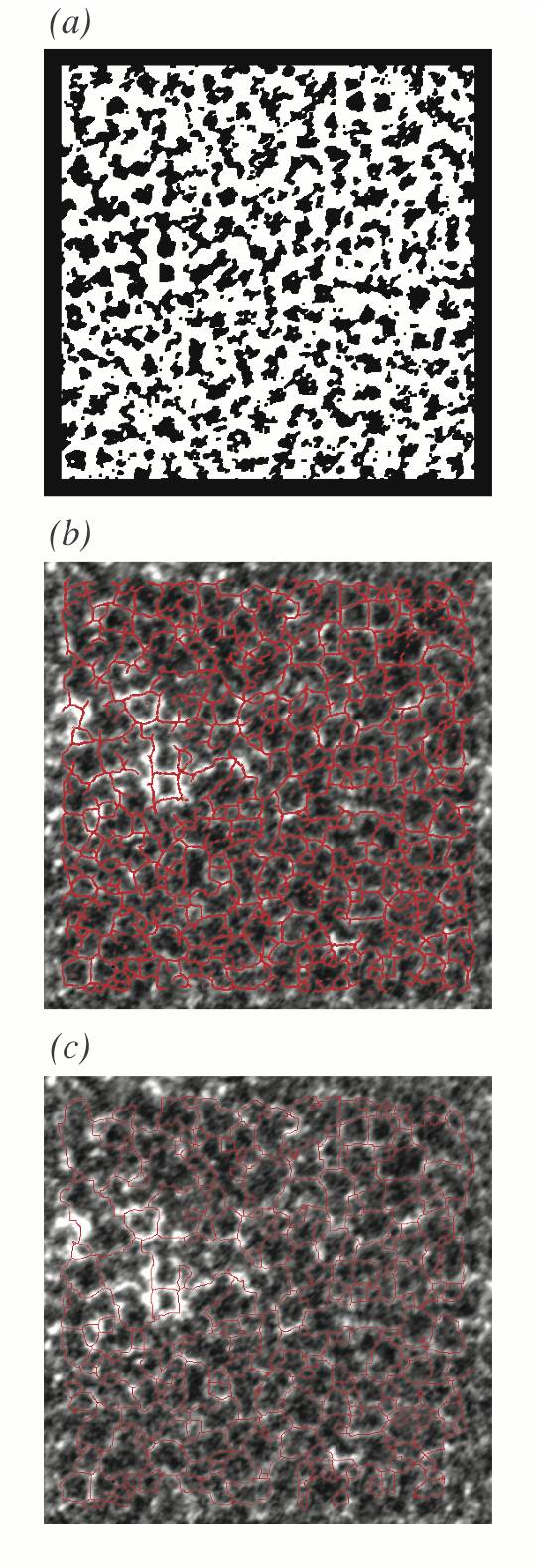}
\end{center}
\caption{(Online) Three steps in the skeletonization process:  The binary image ($a$) is skeletonized, yielding an initial approximation for the supergranular network. This initial skeleton is dilated ($b$) to include locally bright pixels that are connected to network. The resulting dilated network is then reskeletonized.  Dialation and reskeletonization are repeated until differences between successive iterations become neglegible \citep{ber05}. The resulting skeleton is closely aligned with the brightest intensity contours in the image.  The image displayed was recorded at 18:50UT on 7 March 2005.}
\label{skeletonization_fig}
\end{figure}

\begin{figure}
\centering
\includegraphics[scale=.8]{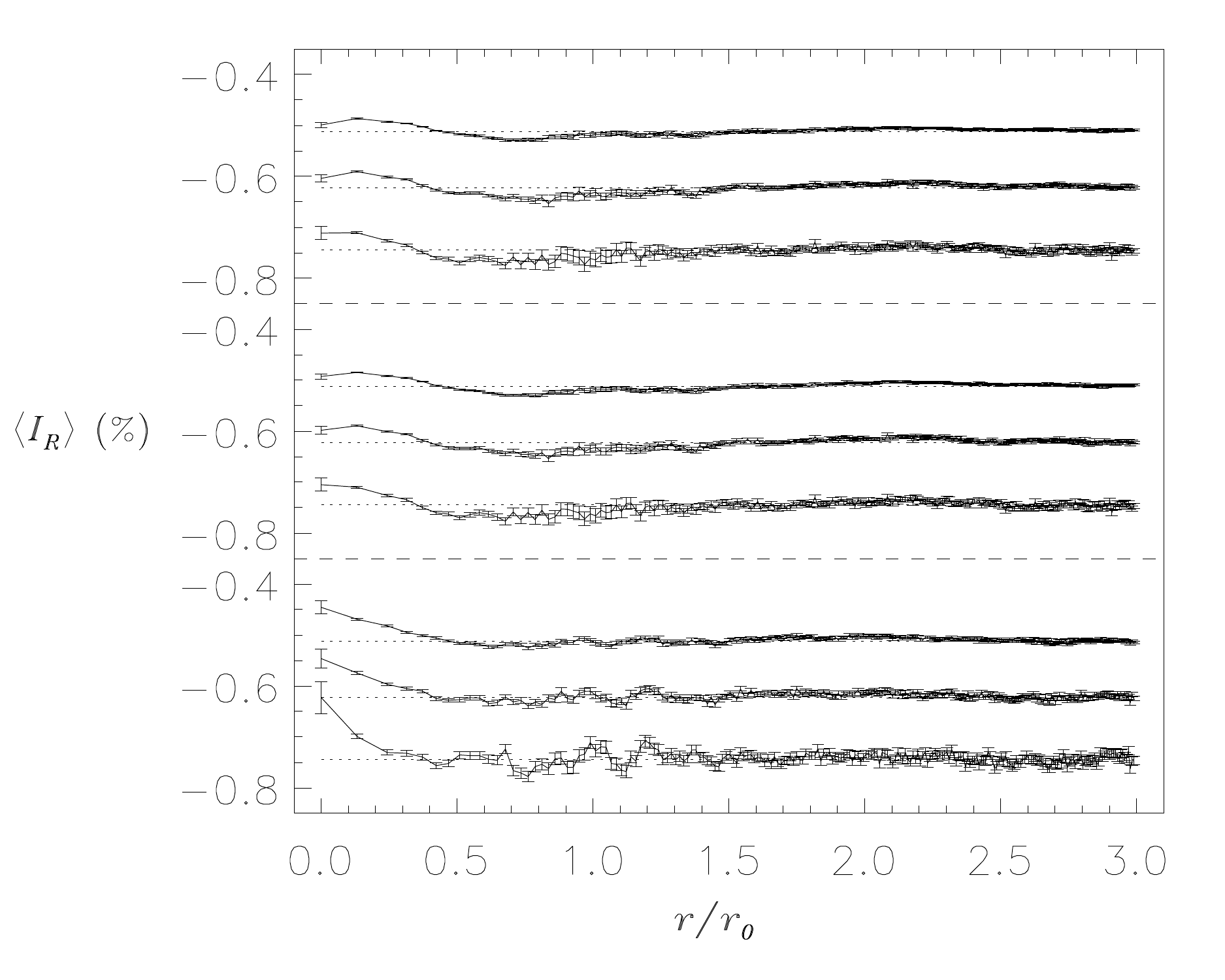}
\caption{(Online) The average supergranular contrast profiles $I_R(r/r_0)$ for masks M5, M6, and M7 and identification parameters, $\xi = -0.6$ with cells smaller than 7.25 Mm excluded (top), $\xi = -0.6$ with cells smaller than 2.2 Mm excluded (middle), and $\xi = -0.2$ with cells smaller than 7.25 Mm excluded (bottom).  {\it Dotted} fiducial lines indicate the average intensity of the unmasked pixels in the images.}
\label{xiplots}
\end{figure}

For this study, we take $L = 40$ pixels ($\sim$ 30 Mm), a typical supergranular length scale, and explore the sensitivity of the measured supergranualar properties to the value of $\xi$ (see Table~1, online material). The parameter range examined follows that leading to optimal skeletonization \citep{ber98,ber05}.  In general, $\xi$ is a negative number and reasonable network-like structures are obtained with $0.0 < \vert\xi\vert < 1.0$.  As $\xi$ approaches zero, the network cells produced by the algorithm tend to grow larger; as
$\xi$ approaches unity, the network is fragmented and cell sizes decrease.  
This is apparent both visually and in the distribution of cell radii, $N(r_0;\xi)$ (Figure~\ref{cellsizehisto}, online material).  
Good, though never perfect, visual agreement between the identified cells and the Ca II K network apparent in the contrast images is obtained using $\xi \approx -0.4$ (see Figure~\ref{skeletonization_fig}, online material).  More importantly, the intensity profiles presented in \S\ref{signature} are robust over a significant range of $\xi$ values (Figure~\ref{xiplots}, online material).

The skeletonization algorithm employed preserves connectivity within the image, making this algorithm ideal for segmenting the network pattern identified in images of moderate spatial resolution but also posing difficulties when that network breaks up into isolated points, as it does for PSPT Ca II K images of highest quality (those taken under the best seeing conditions at MLSO).  To avoid these difficulties, we degrade all images before segmentation by convolution with a Gaussian so that the resulting limb width of the solar disk in the Ca II K image is 6.6 pixels.  These smoothed images are used as inputs for the skeletonization procedure only. Once the supergranules have been identified all further analysis is done at full resolution.  

As discussed in \S~\ref{intro}, the measurement of the radial intensity profile of the supergranulation is complicated by the presence of the magnetic network.  Network flux elements introduce opacity variations which contribute to positive continuum intensity fluctuations at the supergranular boundaries.  The goal is to remove this contribution and examine pixels with only weak integrated magnetic flux density, in order to determine the intensity profile of 'unmagnetized' supergranular plasma.  We note here, and discuss in \S~4, that this separation is likely never complete and is image resolution dependent. 

 \citet{meu07}, employing both continuum images and magnetograms from the Solar and Heliospheric Observatory (SOHO) Michelson Doppler Interferometer (MDI), evaluate several possible strategies for removing or accounting for the magnetic element contribution to the intensity signal.  Intensity correction, based on a measured magnetic-flux continuum-intensity relation, proved inadequate, likely because this relationship has significant scatter due to the unresolved magnetic substructure of the elements \citep{cri09}.  Direct masking \citep{ras03b,ras08} proves more useful, and after image alignment and resizing, we employ that technique here.  The image masks are based on both red continuum and Ca II K intensity which serve as a proxies for the magnetic flux density.  The red continuum image is used to eliminate dark pore, umbral, and penumbral pixels from further consideration.  Bright Ca II K pixels are eliminated with a series of increasingly severe contrast thresholds. As summarized in Table~3 of \citet{ras08}, these masks together eliminate from  0.1 \%  to 99.96 \% of the pixels in a given image from consideration, depending on the severity of the Ca II K threshold.

Continuum intensity fluctuations in the solar photophere are dominated by granulation.  Since the lifetime of a supergranule exceeds that of a granule by a factor of about 100, the intensity fluctuations of individual supergranules should be marginally detectable (equal in magnitude to those of granules) by observations on a subgranular-lifetime cadence if their amplitude is about 10 times smaller than those of granulation. Unresolved quiet-sun granulation displays an average red continuum contrast  of $\sim~1\%$ in PSPT images, making supergranular contrasts of $\sim~0.1\%$ just marginally detectable by image averaging over a supergranular lifetime.  While, as will be shown in \S3, the nonmagnetic supergranular signal is of this order, continuous high quality observation over a supergranular lifetime is not possible from the ground.
This motivates measurements that average over many supergranules, sacrificing knowledge of the individual cell profile for vastly increased signal-to-noise. 

Using the supergranular maps and masks described, we measured the azimuthally averaged supergranular radial intensity profile in the red, blue, and Ca II K contrast images. The mean contrast was determined for each of 225 equal area annuli extending outward from the barycenter of the supergranules to 3.0\,$r_0$, where $r_0$ is average radius of the individual supergranule measured in the four grid-aligned directions.  With this scaling, the outer radius of the 25th annulus corresponds to the measured cell's average radius.  To avoid nested averages, the contrast  values for all of the pixels in any particular annulus were tabulated for all supergranules, and the mean and standard deviation of the contrast in each radial bin were calculated over the entire data set.  The central contrast value for each supergranule was also recorded, and since its average combines only single pixel values from the individual supergranules, the random error in the final profile at  $r = 0$ is larger than at larger radii.  

We note, that by averaging intensities using an annular sampling, we are implicitly assuming that supergranules are circular.  Departures from circularity, evident for most supergranules under consideration, smears the resulting profile, and our final measurement thus represents a lower limit to the contrast variation across a supergranular cell.   Moreover, while equal-area annuli allow measurements of equal significance at all radii for any single supergranule, the measuments have greater variance for small supergranules than large, since the number of radii is fixed and thus the number of pixels in each annuli goes down with supergranular size.   We ameliorate this to some degree by discarding small cells (average radii less than 10 pixels, $\sim 7.3$ Mm) as too poorly sampled, and the results are qualitatively insensitive to the exact threshold (Figure~\ref{xiplots}, online material).  The discarded small cells are about a factor of $2$ smaller than the average supergranule in the images, and belong to the lower portion of a somewhat bimodal size
distribution of radii (Figure~\ref{cellsizehisto}, online material).  They appear to
largely represent substructure (perhaps mesogranulation) within supergranular cells.  

The measurements described were made for all the supergranular cells identified in the images (but not discarded as too small) after application of the eight Ca II K masks, with only unmasked pixels contributing to the analysis.  Since the pixels with the lowest magnetic field strength have on average the lowest contrast, the mean intensity at all radii decreases as the applied mask becomes increasingly severe (Figure~\ref{profiles}).  Of more interest is the radial variation in the contrast profile (cell center to network boundary).  To confirm that that that variation for each mask value is a property of the supergranular cells and not due to unforeseen systematic error, the averaging procedure described was also performed using random pixel locations as cell centers.  Flat profiles, corresponding to the mean intensity of the images after masking ({\it dotted lines} in Figure~\ref{profiles}), were always found for such random sampling.

\section{RESULTS}

\begin{figure}
\centering
\includegraphics[scale=.50]{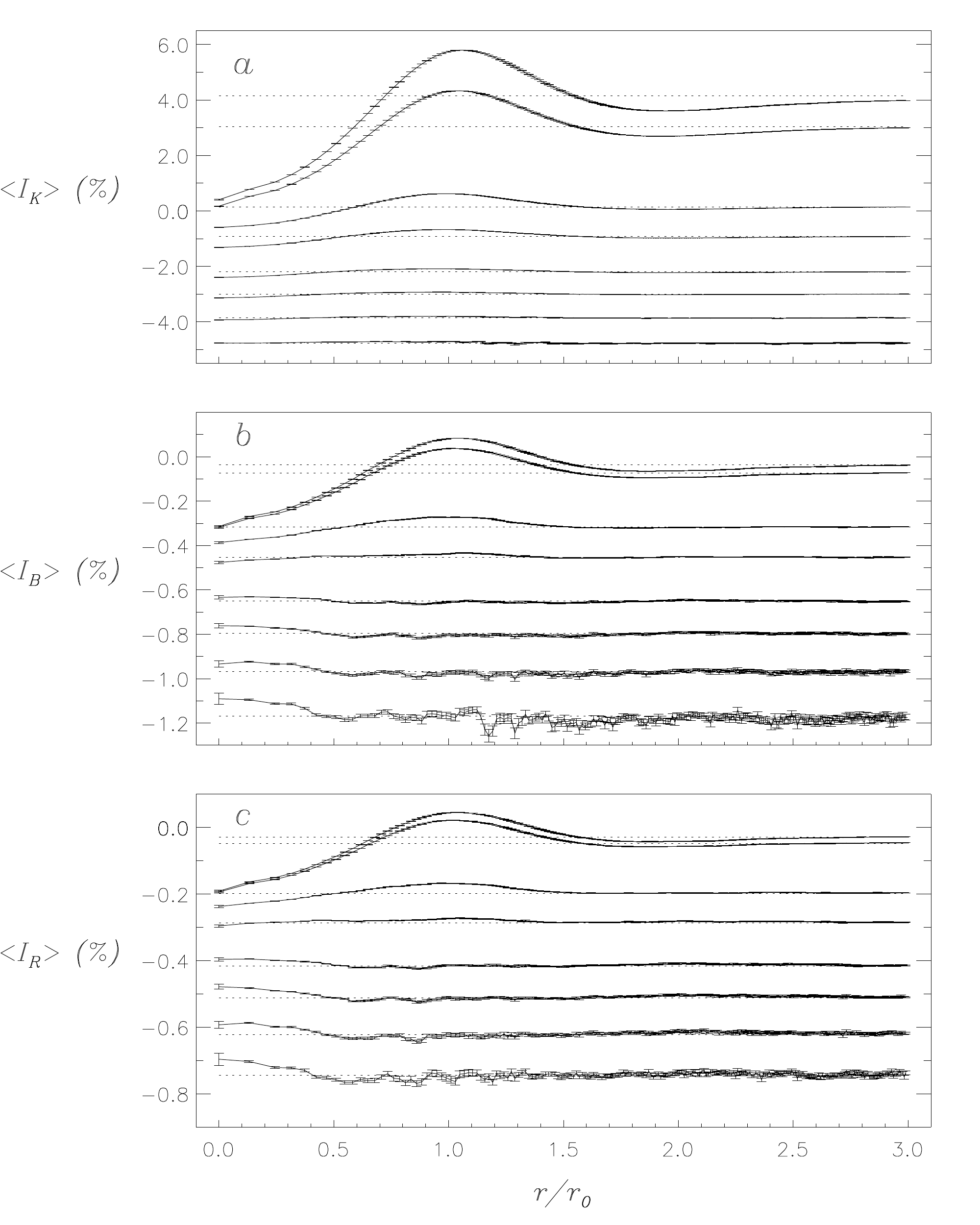}
\caption{The average supergranular contrast profiles $I_c(r/r_0)$ in the ($a$) Ca II K, ($b$) blue continuum, and ($c$) red continuum wavelength bands for each value of the Ca II K masking threshold applied. {\it Dotted} fiducial lines indicate the average intensity of the unmasked pixels in the images. Profiles found using the least severe mask ($\rm M0$) are at top of each plot, with each curve below found using progressively more severe activity masks (mask $\rm M7$ used for the bottommost profile, see Table~3 in \citet{ras08} for mask definitions).  The peak at $r/r_0 = 1$ for weak masking reflects the contribution from magnetic network.  With increasingly severe masking the enhanced intensity at the cell boundaries decreases, eventually dropping, at continuum wavelengths for the most severe mask values, below the intensity observed at the cell centers.  These profiles are for cells identified using the identification parameter $\xi = -0.4$ (Row~2, Table~1 online) and a minimum cell radius of 10 pixels ($\sim7.3$Mm).}
\label{profiles}
\end{figure}

The main results of this study are shown in Figure~\ref{profiles} as plots of the average contrast $<I_c>$ at each wavelength as a function of radial position $r/r_0$.   The results after increasingly severe masking (top to bottom) are  presented as separate curves.   In Ca II K (Figure~\ref{profiles}$a$) the strong increase in brightness associated with the magnetic network in the absence of masking ({\it top most curve}) is seen as a $\sim5 \%$ increase in average contrast at $r/r_o = 1$ above that at cell centers.  The curve approaches the mean value of the image at distances which exceed the correlation length of the network, at which point annular sampling about the cell centers and complete sampling are equivalent.  Notably, the three most severe masks (99.96\% of the brightest pixels masked out for the most severe) applied to Ca II K images produce profiles with a residual contrast difference of only $\sim 0.05 - 0.2 \%$ between cell centers and boundaries, indicating the elimination of nearly all magnetic element contributions, but that residual has the same sense as the unmasked profile, dimmer at cell center and brighter near the boundaries between cells.  

The red and blue continuum images (Figures~\ref{profiles}{\it b} and {\it 4c}) show similar network brightening when no masking is applied, with a  $\sim 0.3-0.4\%$ increase in average contrast from the cell centers to the boundaries, consistent with previous measurements.  This magnetic element contribution to the intensity gradually diminishes as the severity of masking increases.  After application of the more severe masks, the plots show a slight decrease in the average supergranular contrast toward the cell boundaries.  Notably, this is true even though the opposite trend continues to be measured in Ca II K for the same mask values (above).  Moreover, there is some weak indication that, not only does the contrast decrease with distance from the supergranular center, but the cell boundaries are on average slightly darker than the image mean.   

The decrease in contrast from the supergranular cell centers outward after masking the magnetic element contribution is found at both continuum wavelengths, and has a maximum amplitude of $\sim$ 0.05-0.1\%.  The differences between the amplitudes of the signal at the two wavelengths is consistent with a single brightness temperature perturbation (below).  Since more than $10^5$ supergranules are included in the measurement, random errors (or fluctuations since they are largely solar in origin) are small (the one sigma range is over plotted in Fig.~\ref{profiles}) and the statistical significance of the measurement is high.  The results are fairly insensitive to the number of annuli employed. Increasing the number increases the radial resolution, but also the random/granular noise.  Moreover, some inherent smoothing of the profile occurs, independent of the number of anulli employed, because the geometry of the supergranules is not circular as assumed.  Tests with random pixel locations (\S2) suggest that aside from this smoothing, systematic errors in the measurement are even smaller than those due to random fluctuations.  The results found here using $>10^5$ supergranules thus confirm with much greater confidence those of \citet{ras03b}, who tentatively detected the signal using the same techniques applied to 7300 supergranules, and \citet{meu07} who confidently detected such a signature using quite different methodologies on 7629 supergranules.

\begin{figure}
\centering
\includegraphics[scale=.7]{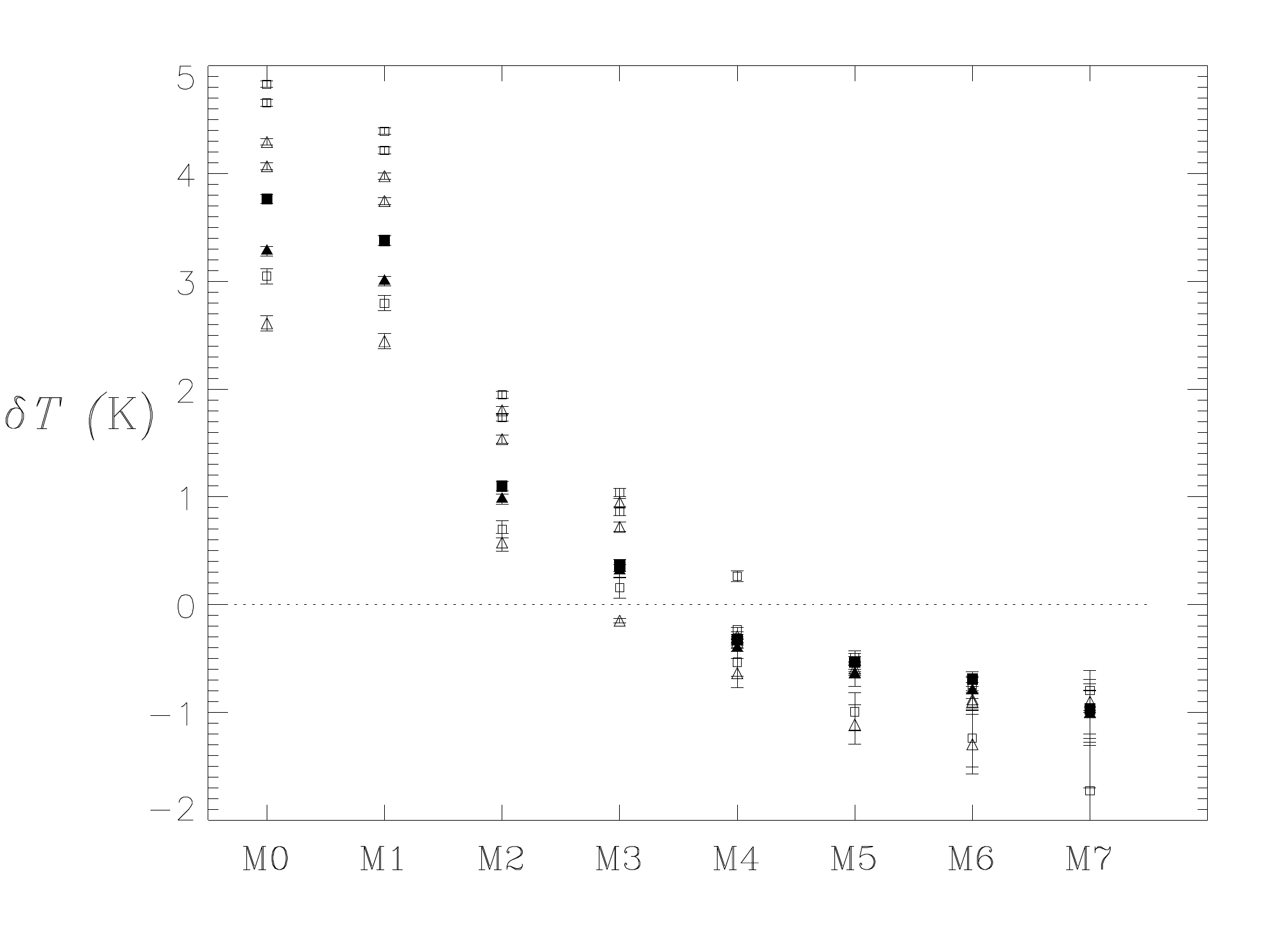}
\caption{{\it Solid symbols} plot the red continuum (triangles) and blue continuum (squares) brightness temperature difference across the supergranular cells as derived from the maxima and minima of the intensity profiles of Figure~\ref{profiles} between $r/r_0=0$ and 1. {\it Dotted} fiducial line  indicates $\delta T=0$.  With no masking, we find an average temperature increase of 3-4 Kelvin from center to edge across the supergranular cells.  For more severe masking (M4 through M7), the brightness temperature difference (edge minus center) becomes negative, converging on a value near $-1$ K.  Open symbols plot the same quantities after varying the identification scheme parameters (see text).}
\label{tempplot}
\end{figure}

Assuming that the observed intensities at the continuum wavelengths are due to small temperature perturbations of a thermal blackbody, the supergranular center-to-edge contrast measured can be converted 
for each mask value into a brightness temperature difference.   These are plotted for the red and blue continuum images with filled symbols in Figure~\ref{tempplot}, where the brightness temperatures were found using the observed maximum and minimum contrast values of the profiles between $r/r_0=0$ and 1 (Figure~\ref{profiles}).  With no masking, the temperature fluctuations associated with the bright magnetic network have amplitudes of $\sim 3-4$ K.   With increasingly severe masking, the fluctuations become negative, with a value near $-1$ K for the most severe masks.  On average, the brightness temperature of the 'nonmagnetized' supergranular plasma is $\sim 1$ K greater at the supergranule center than the edge, consistent with the range 0.8 to 2.8 K cited by \citet{meu07,meu08}.

These results are only weakly sensitive to variation of the identification-scheme parameters.  Contrast profiles resulting from identical analysis of cells identified using three choices of $\xi$ and cutoff size threshold,
are plotted in Figure~\ref{xiplots} (online material).  These curves, for the three most severe masks,

are qualitatively and quantitatively similar to those found using $\xi = -0.4$ and a cutoff of 10 pixels (Figure~\ref{profiles}).  The corresponding average brightness temperature contrast across the cells are plotted as open symbols in Figure~\ref{tempplot}.  The results are all generally within 2 K, with the plot scatter reflecting the sensitivity of the maximum and minimum profile intensities to the identification scheme parameters, and the error bars indicating the formal propagation of random error (largely granular fluctuations) in the intensity measurement.   

\label{signature}

\section{CONCLUSION}

In this work, we have successfully measured the mean supergranular radial intensity profile.  We have employed maps and masks based on the Ca II K intensity of the magnetic network.  After application of the most severe masks, we have found, with a high degree of confidence, that the centers of supergranules are on average $\sim 0.1\%$ brighter than their borders.  This corresponds to a brightness temperature difference of $\sim$ 1 K.  The detection, at the very edge of the PSPT sensitivity threshold, 
is consistent with previous measurements \citep{ras03b,meu07,meu08}, 
this despite the fact that  we employ very different methodologies from Meunier et al. for detection, mapping, and masking of the supergranular cells.

Is the amplitude of the signal observed consistent with a convective origin for the supergranulation?  A simple balancing of the kinetic energy of the motions and buoyancy work over the length of parcel travel \citep{fra70},
 \begin{equation}
{ 1\over 2}\rho v^2\approx\rho^{\prime}{l\over 2}\epsilon\ ,
\end{equation}
 where $\rho$ is the plasma density, $v$ its velocity, $g$ acceleration due to gravity, $l$ the length of parcel travel, and $\epsilon$ a measure of convective efficiency, yields, when evaluated for the linear displacement of an ideal gas parcel in pressure equilibrium, expected temperature perturbations
 \begin{equation}
T^\prime\approx{{Tv^2}\over{gl\epsilon}} ,
\end{equation}
where $T$ is the plasma temperature.  For granulation, $T^\prime$ expected from a strict balance of kinetic energy and buoyancy work (taking $\epsilon=1$) is 33 K and its measured value in the photosphere is $\sim300$K, suggesting that $\epsilon=0.1$.  Numerical models of granular flow suggest that the actual continuum intensity contrast of granulation may in fact be a factor of two yet higher \citep{dan08,nor09}.  The same value for epsilon is deduced for supergranulation, taking $v\sim0.2$km/s, $T\sim10^4$K, $g\sim0.3$km/${\rm s^2}$, $l\sim15$Mm, and the observed temperature contrast measured here $T^\prime\sim1$K.  So while the thermal signal measured is small, both granulation and supergranulation show continuum intensity contrasts about an order of magnitude larger than that expected from the highly simplified parcel argument above but significantly smaller than one might expect from numerical models of granular flows.

A convective origin for the supergranulation may not be an exclusive explanation for the measured intensity perturbations.  We note that the masking procedure employed removes bright magnetic flux elements {\it only in so far as they can be resolved}.  Variation in packing density of unresolved magnetic flux elements may contribute to the brightness profiles measured.  Rather than reflecting a thermal perturbation, the positive continuum intensity signal measured may instead indicate that on average the central regions of supergranular cells contain many small-scale unresolved flux elements that subsequently coalesce as they are advected toward the boundary rendering them resolvable to our masking procedure.  This interpretation is suggested by the apparent ubiquitous presence of inter-network flux elements \citep[e.g.][and references therein]{dew08}, and the degree of masking required to remove the magnetic element signature from the measurement.  Average profiles with enhanced cell center intensities are observed at the red and blue continuum wavelengths only after more than 95\% of the pixels have been masked, leaving only pixels with magnetic flux density $\lesssim 0.7$G (see Table~3 in \citet{ras08} for tabulated properties of the eight masks employed).  The Ca II K intensity, on the other hand, remains enhanced at network cell boundaries even after more severe masking (dark cell centers and brighter boundaries are still seen in Ca II K after masking 99.96\% of the brightest pixels, down to average magnetic flux densities of $\sim0.4$G).  It thus suggests that the number density of unmasked flux elements is less at the cell centers than the network boundary for all mask values.  Three possibilities remain to reconcile these profiles:  1.  the trend in magnetic flux density suggested by the masking sequence on Ca II K images does not continue at unresolved scales and the continuum signal reflects the presence of greater amounts of unresolved flux at cell centers than at the cell boundaries.  2.  an underlying thermal perturbation dominates the contribution of residual flux elements to the continuum, but not the Ca II K, intensities after severe masking (i.e. we have indeed uncovered the convective signature of supergranulation).  3.  the radiative properties of very weak magnetic flux elements or weak field regions, if the field is no longer collected into semi-discrete bundles at those scales, are such that, unlike at greater flux densities, their contrast is positive in Ca II K but negative at both the red and blue continuum wavelengths.  Without detailed modeling of the flow, flux distribution, and radiation field it may not be possible to unambiguously untangle these possibilities.

Finally, we note that the severe masking required to uncover a possible thermal signature of the supergranulation suggests that the concept of 'quiet-sun' may be inappropriate.  Opacities variations due to the presence of magnetic field may be important in determining intensity fluctuations on the Sun at all values and spatial and temporal scales.  The presence of magnetic network may play a important active role in the driving of the supergranular flow by radiatively cooling the deeper layers of the photosphere \citep[e.g.][]{cri09}.  This would imply that the network field is not strictly a passive tracer of the supergranular flow, but rather may play an active dynamic role in supergranular scale selection by inducing radiative perturbations.  This may provide an explanation for why strictly hydrodynamic simulations have difficulty unambiguously obtaining supergranular scales \citep{ust08, ste09}, though domain sizes are also currently still quite restrictive.

\acknowledgments

Special thanks to S. Criscuoli, J. Harder, and an anonymous referee.

\begin{deluxetable}{ccccccc}
\tabletypesize{\scriptsize}
\tablecaption{Supergranule Population Statistics}
\tablewidth{0pt}
\tablehead{
\colhead{$\xi$} & \colhead{$<r_0>$\tablenotemark{(1)}} & \colhead{$\sigma_{r_0}$} & \colhead{$<A>$\tablenotemark{(2)}} & \colhead{$\sigma_A$} & \colhead{$N$\tablenotemark{(3)}} & \colhead{S\tablenotemark{(4)}} \\
 & \colhead{(Mm)} & \colhead{(Mm)} & \colhead{(Mm$^2$)} & \colhead{(Mm$^2$)} & & \colhead{(Mm)}
\label{SGpoptable}
}
\startdata
-0.2 & 16.2 & 7.7 & 1202 & 1775 & 42374 & 7.3 \\
-0.4 & 14.3 & 5.1 & 776   & 675   & 111635 & 7.3 \\
-0.6 & 17.0 & 4.8 & 1004 & 592   & 179898 & 7.3 \\
-0.6 & 15.4 & 5.5 & 861   & 615   & 219773 & 2.2
\enddata
\tablenotetext{(1)}{Average cell radius}
\tablenotetext{(2)}{Average cell area}
\tablenotetext{(3)}{Total number of cells identified}
\tablenotetext{(4)}{Cell radius threshold}
\end{deluxetable}

\end{document}